# Digital Privacy for Migrants: Exploring Current Research Trends and Future Prospects

**Sarah Tabassum**, University of North Carolina at Charlotte
**Cori Faklaris**, University of North Carolina at Charlotte

## Abstract

This paper explores digital privacy challenges for migrants, analyzing trends from 2013 to 2023. Migrants face heightened risks such as government surveillance and identity theft. Understanding these threats is vital for raising awareness and guiding research towards effective solutions and policies to protect migrant digital privacy.

**Keywords:** migrants, newcomers, inclusive privacy and security, human and societal aspects of privacy and security

## Introduction

Privacy is a fundamental right for every individual, but its significance amplifies for marginalized populations who are often neglected in studies on privacy and security. Among these marginalized groups are migrants, defined as individuals who relocate, often in search of better living conditions, according to the Oxford dictionary [25]. An international migrant, consequently, is someone crossing their country's borders. The global trend of international migration has consistently risen in recent years, both in absolute numbers and as a percentage of the world population. In 2020, approximately 281 million people, constituting around 3.6% of the global population, resided in countries other than their own [18, 36]. This rise is driven by forced migration resulting from armed conflicts, natural disasters, and economic crises [1].

Privacy and security concerns are particularly significant for migrants, who face unique risks such as discrimination, exploitation, and physical harm when their personal information is exposed [12, 33, 35, 37]. Migrants often encounter precarious situations during their journeys and resettlement processes, making them vulnerable to various forms of abuse and injustice [12]. Understanding these risks is crucial to emphasize the potential harms and challenges they face. Our paper explores digital privacy challenges for migrants, analyzing trends from 2013 to 2023. Migrants face heightened risks such as government surveillance and identity theft. Understanding these threats is vital for raising awareness and guiding research towards effective solutions and policies to protect migrant digital privacy.

Through a comprehensive review of existing literature and analysis of key trends, we uncovered the complexities of managing digital identities and privacy concerns in migrant communities. Our research highlights the need for tailored solutions and policies to safeguard the digital privacy rights of migrants, considering their unique circumstances and vulnerabilities. The main takeaway of our paper is the identification of significant digital privacy challenges faced by migrant populations, along with potential avenues for addressing these challenges. By exploring this intersection, we aim to raise awareness of the unique challenges faced by migrants in navigating digital spaces and to provide insights that can inform more effective policies and interventions. Ultimately, our goal is to empower migrants to exercise greater control over their digital identities and privacy, fostering inclusivity and dignity in the digital space.

## Privacy Research for Migrants

In 2022, Sannon and Forte emphasized the critical importance of studying privacy issues for marginalized people due to the disproportionate harms they face when they lose their privacy [31]. For instance, marginalized groups may face increased risks of discrimination, exploitation, or even physical harm when their personal information is exposed. Addressing the unique privacy-related needs and behaviors of marginalized individuals is for developing equitable and privacy-protective technologies.

Migrants play diverse roles in the economic and social landscape, influencing local demands, generating indirect financial flows, impacting human capital and knowledge diffusion, and reshaping social norms [3, 4, 27, 33]. Technology, specifically Information and Communications Technologies (ICTs), is instrumental in the mobility process, as migrants rely on it for real-time information and advice before, during, and after their migration journeys [27, 33, 36]. This reliance has sparked interest and occasional concern [33, 36].

Understanding the different categories of migrants is essential when discussing their privacy and security concerns. According to the International Organization for Migration (IOM), classifications such as immigrant, migrant worker, refugee, asylum seeker, and undocumented migrant highlight the varied rights and protections these individuals may have. For example, a **migrant worker** might have limited rights compared to an **immigrant**, who typically enjoys most rights in the host country. A **refugee** has fewer rights but cannot be deported as long as their home country remains unsafe. **Asylum seekers** have fewer rights than refugees and can be deported if their claims are denied, while **undocumented migrants** usually have no political rights or legal status and face significant challenges due to strict migration policies.

Despite researchers working on HCI and social aspects for migrants, few studies have explored the privacy concerns and solutions for this population. According to Sannon and Forte, from 2010-2020 only 8.8% of work within the privacy and security domain for marginalized groups was related to immigrants or migrants [31]. Other marginalized groups included people with disabilities, women in patriarchal societies, victims of intimate partner violence, communities and neighborhoods, and marginalized individuals in general.

The diagram in Figure 1 illustrates the comparison between the population of migrants and other marginalized groups in published papers within privacy and security venues from 2010 to 2020.

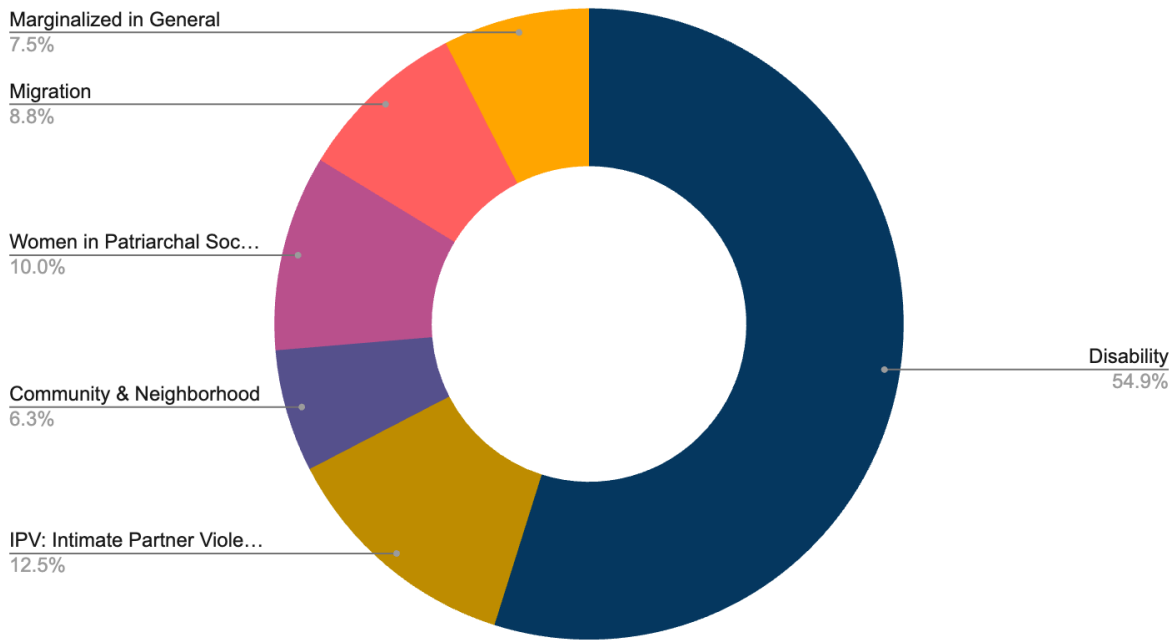

**Figure 1: Comparison of Privacy & Security Venue Publication Percentages: Migrants vs. Other Marginalized Groups (2010-2020). This diagram was created by the first author based on the data shared by Sannon and Forte [31].**

The diagram clearly shows that even within this marginalized group, not much work has been done for migrants. Therefore, with our work, we aim to identify recent trends in privacy research for migrants and discuss future aspects to encourage more privacy research on this population.

## Methodology

In this research, we focused on discerning current trends in the field by comprehensively reviewing papers spanning the period from 2013 to 2023, specifically addressing the intersection of digital privacy and migrant experiences. For the collection of data, we conducted searches using keywords such as "migrants and privacy", "newcomers' privacy", "refugee privacy", "older migrants' privacy", and "digital privacy and migrant'". We explored multiple digital libraries, including ACM Digital Library, USENIX Security, Taylor & Francis Online, Springer, IEEE Digital library, UN iLibrary , and HeinOnline.

Inclusion Criteria involved selecting papers published between 2013 and 2023 that addressed digital privacy issues specifically related to migrants, newcomers, refugees, and older migrants. We included peer-reviewed journal articles, conference papers, and relevant reports from reputable organizations. Our focus was on studies discussing the impact of digital privacy on various categories of migrants, including immigrants, migrant workers, refugees, asylum seekers, and undocumented migrants. Exclusion Criteria involved omitting papers not available in English and articles focusing solely on general digital privacy issues without specific reference to migrants. We also excluded publications that did not provide empirical data or substantial theoretical discussion relevant to digital privacy and migrant experiences. Throughout this investigation, our primary objective was to identify prevalent themes and existing research gaps in the domain. We categorized these themes by engaging in a thorough review, summarization, and reliance on key takeaways extracted from the reviewed papers. This process allowed us to not only comprehend the current state of research but also pinpoint areas where further exploration and emphasis are needed.

Limitations of our search process included the potential inadvertent exclusion of relevant papers due to keyword selection or database limitations. Our focus on specific keywords might have overlooked studies using alternative terminologies for similar issues. Additionally, the exclusion of non-English papers may have limited the scope of our review, potentially omitting valuable insights from non-English-speaking regions.

By shedding light on the dynamics of digital privacy concerning migrants, our work contributes to a nuanced understanding of this crucial intersection. This understanding is enhanced by recognizing the varied rights and legal statuses of different migrant categories, such as immigrants, migrant workers, refugees, asylum seekers, and undocumented migrants, and how these statuses impact their privacy and security needs. For instance, refugees and asylum seekers may face heightened risks of discrimination and exploitation, necessitating robust privacy protections.

## Findings

Through our research, we identified seventeen papers addressing digital privacy and security concerns among migrants. The table below presents the list of these papers, including their publishing venue and year:

**Table 1. The Analyzed Papers**

| | |
|---|---|
| [P1] | Adkins, Denice, and Heather Moulaison Sandy. "Information behavior and ICT use of Latina immigrants to the US Midwest." Information Processing & Management 57.3 (2020): 102072. |
| [P2] | Afnan, T., Zou, Y., Mustafa, M., Naseem, M., & Schaub, F. (2022). Aunties, Strangers, and the {FBI}: Online Privacy Concerns and Experiences of {Muslim-American} Women. In Eighteenth Symposium on Usable Privacy and Security (SOUPS 2022) (pp. 387-406). |

| [P39] | Warford, Noel, Tara Matthews, Kaitlyn Yang, Omer Akgul, Sunny Consolvo, Patrick Gage Kelley, Nathan Malkin, Michelle L. Mazurek, Manya Sleeper, and Kurt Thomas. "SoK: A framework for unifying at-risk user research." In 2022 IEEE Symposium on Security and Privacy (SP), pp. 2344-2360. IEEE, 2022. |
|---|---|

In addition to these papers, several other insightful pieces of literature have contributed to our understanding of privacy perceptions across diverse cultures and ideologies [4, 11, 21, 28]. Particularly noteworthy is the work by Schlesinger et al. [4], which explored the complexities surrounding the disclosure of migrant identity. We have included this paper due to its significant relevance to the discourse on identity disclosure and its consequential impact on privacy considerations.

**Identified Themes in Current Tends**

By reviewing and summarizing the referenced papers and consolidating their key takeaways, we have identified six themes in recent trends of privacy-related research for migrants. The identified themes are: *Social Media and Digital Connectivity, Information Behavior and ICT Use, Privacy Perceptions of Refugees, Asylum Seekers and Undocumented Migrants, Humanitarian Work and Policies, Privacy and Security Vulnerabilities, Mobile Phone Use and Migration*. Table 2 provides an overview of these themes, aligning them with the respective papers within that theme with serial numbers based on Table 1.

**Table 2. Overview of Privacy-Related Research Themes for Migrants**

| Themes | Related Papers |
|---|---|
| **Theme 1:** Social Media and Digital Connectivity | [P3], [P6], [P7], [P11],[P14], [P15], [P18], [P23], [P33] |
| **Theme 2:** Information Behavior and ICT Use | [P1], [P8], [P9], [P19], [P20], [P24], [P28], [P32] |
| **Theme 3:** Privacy Perceptions of Refugees, Asylum Seekers and Undocumented Migrants | [P29], [P31], [P34], [P36], [P37], [P38] |
| **Theme 4:** Humanitarian Work and Policies | [P12], [P13], [P17], [P22], [P27], [P35], [P38], [P39] |
| **Theme 5:** Privacy and Security Vulnerabilities | [P2], [P10], [P16], [P21], [P26], [P29], [P30] |
| **Theme 6:** Mobile Phone Use and Migration | [P4], [P24], [P31] |

The following section discusses each theme, exploring their significant findings in privacy and security research.

**Theme 1: Social Media and Digital Connectivity**

The convergence of social media and digital connectivity significantly influences migrants' experiences and integration. Researchers have explored this area, revealing both potential benefits and challenges. Among them, Cesare et al. (2018) discussed the potential and pitfalls of using social media data for demographic research, highlighting privacy concerns and ethical implications [5]. Similarly, Dekker et al. (2018) examined how Syrian asylum seekers utilize social media for migration decisions, emphasizing its role in information gathering and decision-making while cautioning against misinformation [7]. Gillespie et al. (2018) explored smartphones' dual role in migrants' journeys, serving as tools for navigation and communication but also as potential surveillance instruments [9]. Kim et al. (2022) analyzed privacy risks for migrants in digital communities, stressing the need for enhanced protections [14]. Leurs and Smets (2018) underscored digital tools' potential support for migrants but warned of accompanying risks [17, 18]. Pyle et al. (2024) investigate migrants' perceptions of social media algorithms, revealing concerns about privacy and bias [27]. Monteiro (2022) highlighted social media's impact on migrants' employment outcomes, noting both benefits and digital disparities [22]. These studies underscore the transformative potential of social media in enhancing migrants' lives and facilitating their integration. However, they emphasize the need for policies addressing privacy issues, combating misinformation, and mitigating digital inequalities to ensure social media effectively supports migrants without compromising their rights and safety.

**Theme 2: Information Behavior and ICT Use**

Understanding information behavior and the use of ICT is crucial for migrants' adaptation and integration. Researchers like Adkins, Hsiao, Suh, and Lingel have explored this area extensively. Adkins et al. (2020) highlighted the information-seeking patterns and digital interactions of Latina immigrants in the US Midwest [2], while Newell et al. (2016) focused on information seeking and technology use among migrants at the US-Mexico border [23], emphasizing vulnerabilities due to limited access to reliable information and digital resources. Similarly, Hsiao and Dillahunt (2018) explored how technology can facilitate immigrant access to social capital and aid in their adaptation to a new country, emphasizing the role of ICT in fostering connections and support networks [10]. Suh and Hsieh (2019) investigated Korean immigrants' information behavior and ICT usage during settlement in the United States. They uncovered unique patterns influenced by cultural and linguistic factors [37]. Lingel et al.(2014) and Aricat (2015) examined transnational and low-skilled migrants' online identity work, exploring how they navigate digital spaces to construct and negotiate their identities across borders[12, 29]. Shankar (2023) looked into newcomer communities' information practices and the interdependence between caring for these communities and safeguarding their data, highlighting the complexities of information sharing and privacy in migrant contexts [33]. Together, these studies emphasize the diverse patterns in information behavior and ICT usage among migrants, highlighting the multifaceted nature of their experiences.

**Theme 3: Privacy Perceptions of Refugees, Asylum Seekers and Undocumented Migrants**

Understanding the privacy challenges confronting asylum seekers and undocumented migrants is vital for protecting their rights. Steinbrink et al. (2021) conducted an empirical study on asylum seekers in Germany, shedding light on their digital privacy perceptions during their flight [36]. This work talked about the importance of privacy-enhancing technologies for this vulnerable population. Guberek et al. (2018) and Vannini et al. (2020a) have explored the privacy practices and risks among undocumented immigrants, stressing the need to maintain a low profile in digital spaces to mitigate threats and highlighting procedural challenges and privacy concerns within humanitarian organizations serving undocumented migrants [39, 41]. In another work, Vannini, Gomez, and Newell (2020b) proposed *"Mind the Five"* guidelines for data privacy and security in humanitarian work, prioritizing migrants' privacy and safety [42]. These studies underscore the nuanced privacy challenges faced by comparatively vulnerable migrants, advocating for targeted interventions and policy measures to uphold their rights in digital environments.

**Theme 4: Humanitarian Work and Policies**

Humanitarian work with migrants requires a multifaceted approach that addresses legal, ethical, and technological challenges while safeguarding migrants' rights and dignity. Researchers from sociology and human-computer interaction have explored this critical issue extensively. Among them, La Fors-Owczynik (2016) studied Dutch identity management practices. Their work emphasized the conflict between ensuring security through monitoring and safeguarding migrants' privacy rights [15].The significance of emerging digital infrastructures in border control was highlighted by Latonero and Kift (2018), who emphasized accountability and human rights concerns [16]. Similarly, Milanovic (2015) stressed the need to align surveillance practices with human rights standards to protect migrants' privacy [20]. Another work by Popescu et al. (2022) focused on challenges and solutions for asylum seekers in higher education, advocating for inclusive policies [26]. Vannini, Gomez, and Newell (2019) provide privacy guidelines for humanitarian work with undocumented migrants, emphasizing ethical data handling [32]. Tran et al. (2023) explore security and privacy trade-offs in migration processes, highlighting the need for balanced approaches [40]. Finally, Warford et al. (2022) propose a framework for at-risk user research, informing policies and interventions for vulnerable populations, including migrants [44]. These studies collectively underscore the importance of ethical, rights-based approaches in humanitarian efforts concerning migrants.

**Theme 5: Privacy and Security Vulnerabilities**

Migrants and minority groups contend with heightened privacy and security vulnerabilities [25, 31], necessitating a closer examination of their experiences and challenges. Previous research has focused on the experiences of Muslim-American women and migrant domestic workers (MDWs) in navigating digital landscapes, shedding light on their multifaceted vulnerabilities [3, 13]. Muslim-American women, grappling with complex identities encompassing religious affiliation, gender, immigration status, and race, confront distinct online privacy risks, including Islamophobic harassment and reputational harm within their cultural communities [3]. Similarly, MDWs, often marginalized as passive subjects in digital privacy research, are revealed through participatory workshops to face significant threats such as government surveillance and employer monitoring [13]. McAuliffe et al. (2019) reflect on migrants' contributions amidst increasing disruption and disinformation, highlighting the evolving risks they encounter in digital spaces [19]. These findings underscore the necessity for inclusive security research, recognizing broader social structures contributing to insecurity, and emphasize the importance of tailored privacy strategies, including the establishment of safe online spaces and utilization of community networks. This underscores the significance of prioritizing the digital privacy needs of marginalized communities to foster more equitable and secure online environments.

**Theme 6: Mobile Phone Use and Migration**

Mobile phones play a pivotal role in the migration experiences of individuals, shaping their access to information, communication, and support networks. There have been 3 significant works that specially focus on the usage of mobile for migrants. the significant role of mobile phones in the lives of newcomers, particularly migrants, as they establish themselves in a new country [6, 29, 36]. Coles-Kemp (2018) et al. investigated the impact of mobile phones on migrants' experiences, highlighting how these devices can amplify pressures and reduce capabilities in navigating new environments. Mobile phones serve as a source of security and a means to create a safe space, allowing individuals to maintain connections with their past while navigating their new environment [6, 29]. However, this reliance on mobile technology also brings forth a range of threats and vulnerabilities, extending beyond conventional security concerns. Among low-skilled migrants in Singapore, mobile phone usage was found to enhance their functioning, yet hierarchical structures and overdependence posed constraints to their capabilities [29]. Similarly, asylum seekers in Europe utilize smartphones for information and communication during their flight, but are acutely aware of privacy risks such as surveillance and persecution [36]. This underscores the importance of understanding migrants' digital privacy perceptions and behaviors. The findings call for policies and support strategies that address the specific needs for digital privacy protection among migrants, emphasizing the design of assistance apps and collaboration platforms tailored to safeguarding their privacy and trust in digital environments.

# Discussion and Future Work

The literature review and analysis of the discussed papers clearly indicate that there is much more to be explored in this field. Our study highlights the complexities of digital privacy issues for migrants and emphasizes the need for further research to address emerging challenges and gaps in knowledge. By identifying key research priorities and outlining potential avenues for exploration, we aim to foster a more comprehensive understanding of digital privacy issues for migrants and facilitate the development of effective strategies and interventions to safeguard their rights and well-being. This combined Discussion and Future Work section provides a roadmap for future research endeavors:

## Double Presence: Navigating Online and Offline Identities

The concept of "double presence" presents significant challenges to the privacy of migrants [6], encompassing both online and offline aspects of their lives. Understanding how migrants manage these dual identities is crucial for addressing privacy considerations effectively. Future research should dive deeper into the complexities of maintaining connections with both home and host communities, particularly among educational migrants with distinct mobile-centric lifestyles.

## Addressing Privacy Concerns in Developing Regions

Privacy concerns in developing regions, especially in the Global South and Africa, represent a critical area for future research. Cultural nuances, knowledge gaps, and contextual factors influence diverse privacy preferences in these regions. Exploring culturally embedded privacy practices and developing inclusive privacy policies tailored to local contexts can offer innovative solutions to privacy challenges.

## Long-Term Settlement and Multi-Dimensional Privacy Considerations

The long-term settlement process of migrants entails a multi-dimensional understanding of cultural adaptation, family dynamics, and day-to-day challenges. Future research should examine privacy and security considerations during this intricate process, including intergenerational issues faced by children born and raised in new communities.

## Privacy Concerns of Older Migrants

Privacy concerns of older migrants in new environments, particularly in navigating digital connectivity challenges during old age, warrant further investigation. Research should focus on designing user-friendly interfaces to facilitate continued connections with family members abroad while addressing the unique needs of older migrants.

**Leveraging Generative Artificial Intelligence for Digital Inclusion**

Exploring the role of generative artificial intelligence (AI) in mitigating the digital divide presents an intriguing avenue for future research. Investigating how generative AI can address disparities in access and digital literacy, while considering privacy perceptions among marginalized populations, is essential for ensuring equitable access and safeguarding the security of migrant communities.

## Conclusion

Our work offers a comprehensive exploration of privacy issues concerning migrants, highlighting the critical need for in-depth investigation and innovative solutions tailored to their unique needs. It sheds light on the evolving landscape of international migration and the integral role played by migrants, particularly in integrating digital privacy into their journeys. The literature review underscores gaps in existing studies, emphasizing the necessity for further exploration into diverse aspects of migrants' privacy experiences. Our identification of six key themes in recent trends provides a foundation for future research directions. The outlined future prospects, including exploring the "double presence" issue, designing for privacy in developing regions, studying long-term settlement processes, focusing on older individuals' privacy concerns, and using generative AI for digital inclusion, serve as a roadmap for researchers and practitioners. Through these efforts, we aim to contribute to a more comprehensive understanding of privacy dynamics among migrants, fostering equitable technologies and ensuring a secure digital future for all.